\documentclass[
amsmath, %
floatfix, %
twocolumn, %
reprint, %
prl, %
aps, %
]{revtex4-1}
\renewcommand{\thispagestyle}[1]{}
\usepackage[T1]{fontenc}
\usepackage[utf8]{inputenc}
\usepackage{graphicx}
\usepackage{latexsym}
\usepackage{amsthm}

\usepackage{color}
\usepackage{mathtools}
\usepackage{xcolor}
\usepackage[low-sup]{subdepth}

\usepackage[]{newtxtext}
\usepackage[subscriptcorrection,nosymbolsc,smallerops,bigdelims]{newtxmath}
\DeclareMathAlphabet{\mathcal}{OMS}{cmsy}{m}{n}

\DeclareMathAlphabet{\mathcalb}{OMS}{cmsy}{b}{n}
\usepackage{bm}
\usepackage{hyperref}
\hypersetup{
	colorlinks,
	linkcolor={blue!90!black},
	citecolor={blue!90!black},
	urlcolor=	{blue!90!black}
}
\makeatletter
\renewcommand*{\eqref}[1]{%
	\hyperref[{#1}]{\textup{\tagform@{\ref*{#1}}}}%
}
\makeatother

\DeclarePairedDelimiter\lr{\lparen}{\rparen}
\DeclarePairedDelimiter\Lr{\lbrack}{\rbrack}
\DeclarePairedDelimiter\LR{\lbrace}{\rbrace}
\DeclarePairedDelimiter\abs{\lvert}{\rvert}
\DeclarePairedDelimiter\avg{\langle}{\rangle}

\DeclarePairedDelimiterX{\comm}[2]{\lbrack}{\rbrack}{#1, #2}
\DeclarePairedDelimiterX{\acomm}[2]{\lbrace}{\rbrace}{#1, #2}

\DeclarePairedDelimiter\ket{\lvert}{\rangle}

\DeclarePairedDelimiterX{\braket}[2]{\langle}{\rangle}{#1\delimsize\vert #2}
\DeclarePairedDelimiterX{\matrixel}[3]{\langle}{\rangle}{#1 \delimsize\rvert #2 \delimsize\lvert #3}

\newcommand{\flatfrac}[2]{\lr{#1/#2}}
\renewcommand{\tilde}[1]{\widetilde{#1}}
\newcommand{\upcapt}{\vspace{-0.5em}}

\mathchardef\mhyp="2D

\begin{document}
\title{Quantum glass of interacting bosons with off-diagonal disorder}
\author{A. M. Piekarska}
\author{T. K. Kope\'{c}}
\affiliation{Institute of Low Temperature and Structure Research, Polish Academy of Sciences, 
PO.\ Box 1410, 50-950 Wroc\l{}aw 2, Poland}
\pacs{05.30.Jp, 75.10.Nr, 67.85.Hj, 05.30.Rt}
\begin{abstract}
We study disordered interacting bosons described by the Bose-Hubbard model with Gaussian-distributed random tunneling amplitudes. 
It is shown that the off-diagonal disorder induces a spin-glass-like ground state, characterized by randomly frozen quantum-mechanical 
U(1) phases of bosons. To access criticality, we employ the ``$n$-replica trick", as in the spin-glass theory, and the Trotter-Suzuki method 
for decomposition of the statistical density operator, along with numerical calculations. The interplay between disorder, 
quantum and thermal fluctuations leads to phase diagrams exhibiting a glassy state of bosons, which are studied
as a function of model parameters. The considered system may be relevant for quantum simulators of optical-lattice bosons,
 where the randomness can be introduced in a controlled way. The latter is supported by a proposition of experimental realization of the system in question.
\end{abstract}
\maketitle
{\it Introduction.}--- Understanding the effects of randomness combined with interactions is a major challenge
in condensed matter physics \cite{balian}.
Especially, quantum phase transitions in disordered systems are different in nature in many
aspects from their classical counterparts \cite{sachdev}.
In this context, ultracold bosonic atoms in optical lattices \cite{zoller} represent an
extremely powerful tool for engineering quantum systems with a broad
tunability of parameters, thus serving as quantum simulators \cite{Feynman1982}.
A natural extension of these experiments is the realization of disordered systems
using ultracold atoms in optical potentials \cite{lewen},
as documented by the experimental observation of Anderson localization of matter
waves in a random potential \cite{roati,billy}.
It is now well established that random on-site ({\it i.e.} diagonal) disorder
can destroy the direct superfluid to Mott
insulator transition via the so-called {\it Bose glass}
phase \cite{fisher} in the strongly-interacting limit,
whose characterization has been the object of a number of theoretical \cite{gimp}
and experimental \cite{falani} investigations.
While the effects of diagonal
disorder have been widely recognized, studies of random hopping 
amplitudes, belonging to the 
{\it off-diagonal} category, are scarce (for Monte Carlo simulations in
one-dimensional models, see \cite{prokof,balab,sengupta}).
The importance of this kind of disorder in bosonic systems lies in the fact that it allows one to make contact
with interesting and unexplored features from the realm of spin-glasses (SG) \cite{koster,domini}. The latter
has long left the regime of random
classical magnetic materials and splashed down into many other
areas as neural networks \cite{amit}, high-$T_\mathrm{c}$ superconductivity \cite{auer}, or quantum chaos \cite{hake}.
Two main ingredients are singled out as crucial to set the physical behavior 
of these systems: strong interaction {\it and} frustration.
Because of frustration, the ground state is 
degenerate and often separated by
macroscopically large energy barriers forcing the system to get trapped,
depending on its history, in one of its degenerate local minima.
However, in an interacting disorder-frustrated system a boson, being a quantum object, may
not necessarily be trapped by barrier height since it may be able to tunnel through such barriers,
provided the integrated tunneling probability is finite. This suggests
an interesting competition between frustration and quantum effects as it is manifested
in the {\it quantum spin glasses} \cite{bray}.
There has been an intense interest in studying these systems regarding the nature of the ordered phases and the
transition between them in various contexts \cite{usadel,kopec,buttner,blinc,jose,doman}.
In fact, due to the tunability of quantum fluctuations of the interacting lattice bosons, the nature of various quantum phases in many-body systems with random and frustrated interactions deserves a detailed study.

Here, we address these issues by studying the impact of
random hopping amplitudes on the quantum states in an interacting bosonic system.
The main target is to determine the glassy-phase 
threshold, which can be expressed through basic thermodynamic parameters of interacting bosons subject to disorder.

{\it Model.}---
We model our system with the Bose-Hubbard Hamiltonian \cite{fisher},
\begin{equation}
		\widehat{H} = -\sum_{ij} \lr[\big]{ J_{ij}\hat{a}_i^\dagger \hat{a}_j + \delta_{ij} \mu \hat{n}_i }
+ \frac{U}{2}\sum_i \hat{n}_i\left(\hat{n}_i-1\right) 
		 \label{ham}
\end{equation}
where $\hat{a}_i$ ($\hat{a}_i^\dagger$) are the annihilation (creation) operators for site $i$ ($i=1\dots N$; $N$ is the number of sites) and $\hat{n}_i=\hat{a}_i^\dagger \hat{a}_i$ represent the particle number operators. Furthermore, $\mu$ and $U$ denote the chemical potential and on-site interaction strength, respectively, while $J_{ij}= J_{ji}$, $(J_{ii}\equiv 0)$ stand for independent random variables describing the hopping energies between sites $i$ and $j$. In the following, we assume identical zero-mean Gaussian distribution of all $J_{ij}$. To get a sensible thermodynamic limit, one has to scale the variance of the distribution by $N$, {\it i.e.}, it is given by $J^2/N$. It makes a close contact with the canonical example of Sherrington-Kirkpatrick (SK) \cite{sk} model, which stands as a reference in the SG theory.
Surprisingly, it also permits an experimental realization, where the full connectivity and disorder can be implemented (see {\it Experimental realization}).

{\it Glass signature.}--- 
Since the long-range order is absent in the glassy state, the superfluid order parameter $\langle \hat{a}_i \rangle$ is not
a good quantity to characterize the new state of matter. In fact, in the glassy state the phases of complex quantum-mechanical wave-functions
of bosons tend to freeze in certain directions that randomly change from site to site.
This is signaled in terms of the Edwards-Anderson (EA) order parameter \cite{ea},
$\cramped{{\cal Q}_{\rm EA}} = \sum _i \cramped{\Lr{ \abs{ \avg{ \hat{a}_i }}^2 }_J}/N$,
where $\avg{\cdots} \equiv {\rm Tr} \cdots \exp\lr{-\beta\widehat{H}}/Z$ is the statistical average
and $Z= {\rm Tr} \exp\lr{-\beta \widehat{H}}$ is the partition function for Hamiltonian \eqref{ham}. 
Since the disorder in the system under study is {\it quenched}, one has to additionally perform the configurational
averaging over the random distribution of $J_{ij}$, denoted by $\Lr{\cdots}_J$.
Note, that the transition to a glassy state with $\cramped{{\cal Q}_{\rm EA}}\neq 0$ 
can be driven by both thermal or {\it quantum} fluctuations. Close to $T=0$ the latter dominate and
the nature of this zero-temperature transition is of great interest.

{\it Methods.}---
We start with evaluation of the free energy averaged over the disorder, ${\cal F}$. To this end, we employ the replica method \cite{sk}, which, although known mainly from the classical regime, has been successfully applied to the quantum spin problem as well ({\it cf}.\ Ref.~\cite{usadel}), in agreement with other methods \cite{Kopec1990}. As we are introducing this approach to a significantly different system, we present main steps of our derivation.
The replica trick is based on the representation ${\cal F} = - \lim_{n\to 0}\flatfrac{1}{\beta n}\lr{ \Lr{Z^n}_J-1 }$, where $Z^n = \mathrm{Tr} \exp\lr{-\beta\sum_{\alpha} \widehat{H}_\alpha }$ stands for the replicated partition function involving Hamiltonians
$\widehat{H}_\alpha$ labeled with the replica index $\alpha=1,\dots, n$,
written as ${\widehat{\cal H}}\equiv\sum_\alpha \widehat{H}_\alpha=\widehat{{\cal H}}_P+\widehat{{\cal H}}_Q+\widehat{{\cal H}}_U$, with
$\widehat{{\cal H}}_U = \flatfrac{U}{2} \sum_{i\alpha}\hat{n}^2_{i\alpha}-\tilde{\mu}\sum_{i\alpha}\hat{n}_{i\alpha}$,
$\widehat{{\cal H}}_X = -\sum_{\alpha, i < j} J_{ij}\widehat{X}_{i\alpha}\widehat{X}_{j\alpha}$,
where $X=P,Q$, $\tilde{\mu} \equiv \mu+{U}/{2}$ and the initial Hamiltonian is rewritten using $\widehat{P} = {i}\lr{ \hat{a}^\dagger\!-\hat{a} }/{\sqrt{2}};~\widehat{Q} = \lr{ \hat{a}^\dagger\!+\hat{a} }/{\sqrt{2}}$. 
	
In the quantum Hamiltonian \eqref{ham}, 
the operators do not commute, which obstructs the handling. To circumvent this
difficulty we resort to the generalized Trotter-Suzuki formula \cite{trotter},
	\begin{equation}
		e^{ -\beta \widehat{{\cal H}}} = \!\lim_{M\to\infty} \lr[\Big]{ \prod_Xe^{-{\beta \widehat{{\cal H}}_X}/{M} } }^M\!,~~X=P,Q,U,
		\label{tsuzuki}
	\end{equation}
which is mathematically rigorous for $M\to\infty$. 
To proceed, we split each pair of consecutive exponents in Eq.~\eqref{tsuzuki} with a sum of projectors onto a complete set of either $\widehat{P}$ or $\widehat{Q}$ eigenstates:
$\widehat{P}_{i\alpha}\ket{p^{(k)}_{i\alpha}}= p_{i\alpha}^{(k)} \ket{p^{(k)}_{i\alpha}};~\widehat{Q}_{i\alpha}\ket{q^{(k)}_{i\alpha}}= q_{i\alpha}^{(k)} \ket{q^{(k)}_{i\alpha}}$. Then
	\begin{eqnarray}
		\mkern-32mu Z^n = \mathrm{Tr}\prod_{k=1}^M
		\exp\cramped{ \Lr[\bigg]{ \frac{\beta}{M}\!\sum_{\alpha,i < j} \!\! J_{ij} \lr[\Big]{ p_{i\alpha}^{(k)}p_{j\alpha}^{(k)}\!+q_{i\alpha}^{(k)}q_{j\alpha}^{(k)} } }}\mathcal{M}\lr{p,q},
		\label{zn}
	\end{eqnarray}
where $\mathcal{M}\lr{p,q} = \prod_{k=1}^M \braket{p^{(k)}}{q^{(k)}} \matrixel{q^{(k)}}{e^{-{\beta \widehat{H}_U}/{M}}}{p^{(k+1)}}$ and $\ket{x^{(k)}} \equiv \bigotimes_{i,\alpha}\ket{x_{i\alpha}^{(k)}}$ for $x=p$, $q$.
The resulting {\it classical} model in Eq.~\eqref{zn} turns out to be anisotropic,
with a correlated disorder along the additional time-like axis, which effectively increases the dimensionality of the system by one.
It comes from the division of the ``imaginary time'' interval $[0,\beta]$, over which the system evolves, into many
$\beta/M$-wide subintervals with $M\in\mathbb{N}^{+}$ (the Trotter number), for which the density matrix has to be calculated.

After performing the Gaussian integrals over $J_{ij}$,
we apply the Hubbard-Stratonovich transformation to the various quartic terms in $[Z^n]_J$, which reduces the further evaluation
to a single-site problem and introduces three sets of auxiliary integration variables $\cramped{\lambda^X}$ $(X=P$, $Q$, $PQ$), so that
$[Z^n]_J= \cramped{\int \Lr{ D\lambda^P D\lambda^Q D\lambda^{PQ} } e^{-N\beta {\mathcal{F}}_\mathrm{eff}}}$, where the effective free-energy function is
\begin{eqnarray}
		{\mathcal{F}}_\mathrm{eff} &=& \beta^{-1} \sum_{k\alpha k'\!\alpha'} \Lr[\Big]{ \lr[\big]{ \lambda_{k\alpha k'\!\alpha'}^P}^2 + \lr[\big]{ \lambda_{k\alpha k'\!\alpha'}^Q}^2 + {\lr[\big]{ \lambda_{k\alpha k'\!\alpha'}^{PQ}}^2}/{2}}\nonumber\\
		&&- \beta^{-1} \ln \mathrm{Tr} \exp \lr{-\beta H_{\rm eff}},
		\label{free}
	\end{eqnarray}
with the effective classical Hamiltonian
	\begin{align}
		H_{\rm eff} = & \frac{1}{N\beta}\LR[\bigg]{
		\sum_{k\alpha}\frac{J\beta}{2M}\Lr[\Big]{\lr[\big]{p_{\alpha}^{(k)}}^2 +\lr[\big]{ q_{\alpha}^{(k)}}^2 } }^2 \nonumber\\
		&- \frac{J}{M}\!\sum_{k\alpha k'\!\alpha'}\left[
		\lambda^P_{k\alpha k'\!\alpha'}p_{\alpha}^{(k)}p_{\alpha'}^{(k')}
		+ \lambda^Q_{k\alpha k'\!\alpha'}q_{\alpha}^{(k)}q_{\alpha'}^{(k')}\right. \nonumber\\
		&- \left.\lambda^{PQ}_{k\alpha k'\!\alpha'}p_{\alpha}^{(k)}q_{\alpha'}^{(k')}\right]
		+ \frac{1}{\beta}\ln \mathcal{M}(p,q).
		\label{effham}
	\end{align}

{\it Self-consistent solution.}---
In the large-$N$ limit, one can evaluate $[Z^n]_J$ exactly, using the
saddle-point method. The corresponding self-consistent equations for $\lambda$-parameters are
	\begin{eqnarray}
		&&\lambda^P_{k\alpha k'\!\alpha'} = \frac{J\beta}{2M}\avg[\big]{p_{\alpha}^{(k)}p_{\alpha'}^{(k')}},\quad\lambda^Q_{k\alpha k'\!\alpha'}
			= \frac{J\beta}{2M}\avg[\big]{q_{\alpha}^{(k)}q_{\alpha'}^{(k')}},\nonumber\\
		&&\lambda^{PQ}_{k\alpha k'\!\alpha'} = \frac{J\beta}{M}\avg[\big]{p_{\alpha}^{(k)}q_{\alpha'}^{(k')}},
		\label{sceq}
	\end{eqnarray}
with $\avg{\cdots}$ performed using $H_{\rm eff}$ from Eq.~\eqref{effham}.
Due to the dynamical Trotter time-like dependence of $\lambda$'s, solving the
mean-field equations~\eqref{sceq} remains a rather difficult task.
In this regard, the computational problem of a quantum glass bears some resemblance to the
dynamical mean-field theory \cite{dmft}, widely employed to study correlated fermionic systems.

The symmetries present in the system imply $\avg{ \cramped{p_{\alpha}^{(k)}p_{\alpha'}^{(k')}} } = \avg{ \cramped{q_{\alpha}^{(k)}q_{\alpha'}^{(k')} }}$ and
$\avg{ \cramped{p_{\alpha}^{(k)}q_{\alpha'}^{(k')} }} = 0$,
which allows us to consider saddle-point parameters 
$\cramped{\lambda_{k\alpha k'\!\alpha'}}\equiv\cramped{\lambda^P_{k\alpha k'\!\alpha'}} = \cramped{\lambda^Q_{k\alpha k'\!\alpha'}}$,
$\cramped{\lambda^{PQ}_{k\alpha k'\!\alpha'}} = 0$.
Regarding the replica structure, these variables are of two types, according to the decomposition
$\lambda_{k\alpha k'\!\alpha'}={\cal R}_{k k'}\delta_{\alpha \alpha'}+\lr{1-\delta_{\alpha \alpha'}}{\cal Q}_{\alpha \alpha'}$.
Terms with $\alpha=\alpha'$, denoted by ${\cal R}_{kk'}$, represent dynamic self-interactions that depend only on the difference $\abs{k-k'}$ due to the time-translational invariance, while those with $\alpha\neq\alpha'$ (${\cal Q}_{\alpha\alpha'}$) are 
purely static and related to the EA order parameter,
${\cal Q}_{\rm EA}=\lim_{n\to 0}[2/n(n-1)]\sum_{\alpha>\alpha'}{\cal Q}_{\alpha\alpha'}$.

To locate the critical lines, we expand the disorder-averaged free energy in Eq.~\eqref{free} with respect to the glass order parameters ${\cal Q}_{\alpha \alpha'}$ in the Landau-theory manner. The equation $\left.{\partial^2\mathcal{F}}/{\partial{\cal Q}_{\alpha\alpha'}^2}\right|_{{\cal Q}=0}\!=0$
gives, by utilizing the time-translational invariance, the condition for the appearance of the glassy phase, $\sum_k {\cal R}_{kk'} = {1}/{2}$.
 
{\it Numerical evaluation.}---
The self-consistent equations~\eqref{sceq} are solved numerically for $M$ up to 12. The bottleneck is the thermal average in the expression for ${\cal R}_{kk'}$. Although the imaginary part of summands may be omitted due to symmetry, a severe sign problem \cite{Troyer05} precludes the application of the Monte Carlo method (MC). To check its severeness, we have computed the number of configurations needed for MC by calculating the average sign of the summand. It was comparable to the total number of configurations, hence no polynomial-time solution exists \cite{Troyer05}.
Thus, a direct summation over all configurations of $\cramped{p^{(k)}}$ and $\cramped{q^{(k)}}$ was performed on a high-performance computing cluster.
Since the eigenbases of $\widehat{P}$ and $\widehat{Q}$ (or $\hat{a}$ and $\cramped{\hat{a}^\dagger}$) are infinite, a truncation of the discrete Hilbert space was needed. From convergence tests, we found that the $n=0,1,2$ basis is sufficient for $0\le \mu/U \le 1$, which gives $\cramped{9^M}$ possible configurations, as compared to $\cramped{2^M}$ in the half-integer quantum SG case \cite{usadel}. To obtain meaningful results for $0\le \mu/U \le 2$, we had to enlarge the basis to $n=0,1,2,3$ (and 4, for calculation of $\avg{\hat{n}}$), at the expense of a substantial rise to $\cramped{16^M}$ ($\cramped{25^M}$) configurations.

\begin{figure}[tbp!]
	\includegraphics[width=\columnwidth]{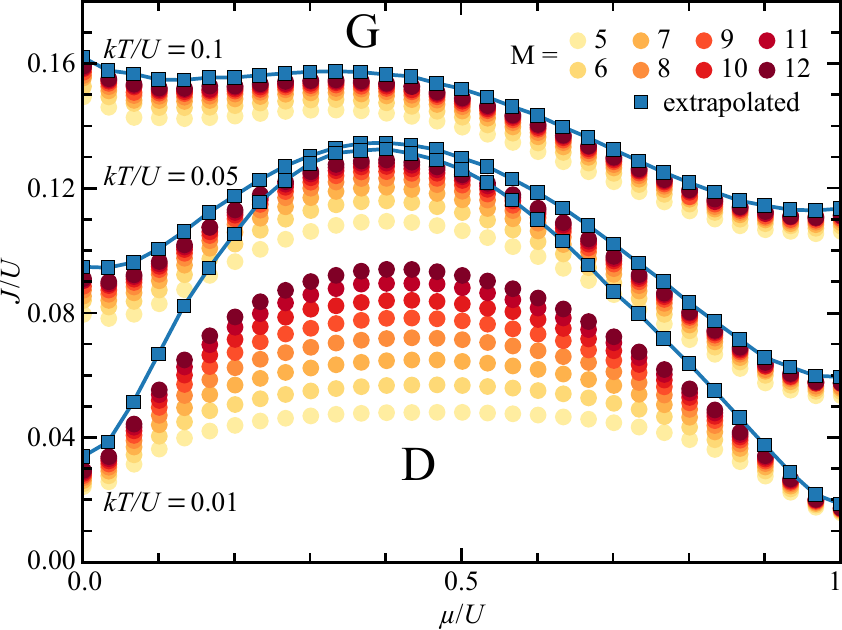}\upcapt
	\caption{Phase diagrams at various temperatures for $M=5\mhyp12$ (circles) and extrapolated to $M = \infty$ (squares). Disordered (D) and glassy (G) phases are marked. Lines are to guide the eye.}
	\label{b3}
\end{figure}
\begin{figure}[tbp!]
	\includegraphics[width=\columnwidth]{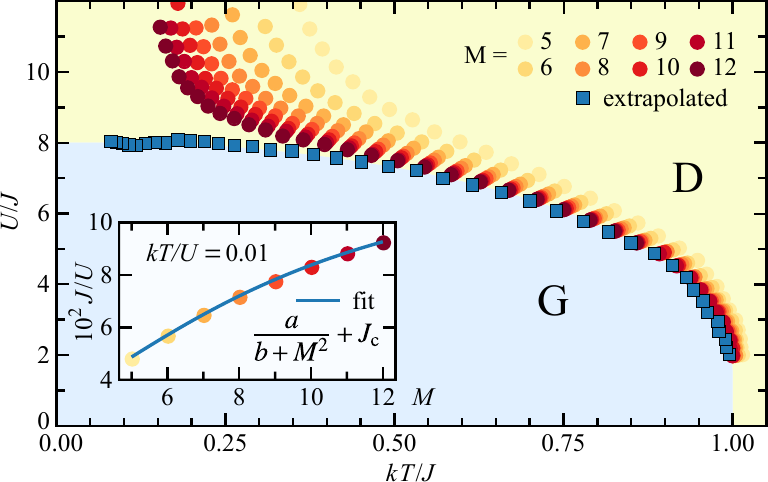}\upcapt
	\caption{Phase diagram in terms of $U$-$kT$ variables at $\mu/U = 0.5$ for $M=5\mhyp12$ (circles) and extrapolated to $M = \infty$ (squares; exemplary fit in the inset). Disordered (D) and glassy (G) phases are marked.}
	\label{temp}
\end{figure}
\begin{figure}[tbp!]
	\includegraphics[width=\columnwidth]{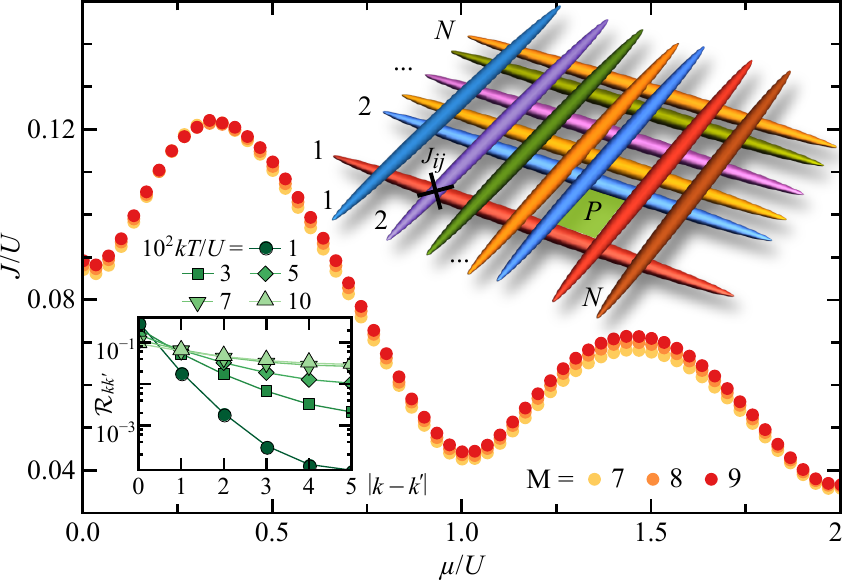}\upcapt
	\caption{Phase diagram at $kT/U = 0.05$ with the Hilbert-space basis enlarged to 4 states, for $M=7\mhyp9$ (circles). Bottom inset: dynamic self-interactions $\mathcal{R}_{kk'}$ for $\mu/U = 0.5$ and $J=J_c$ vs.\ $|k-k'|$ at various temperatures; lines are to guide the eye. Top inset: a schematic view of the proposed experimental realization of considered system; a plaquette $\mathcal{P}$ and tunneling site $J_{\ell\ell'}$ are marked.}
	\label{b4}
\end{figure}	
{\it Results.}---
Solving the self-consistent equations~\eqref{sceq} numerically, we obtain phase diagrams for $M$ up to $12$ at three chosen temperatures and collect them in Fig.~\ref{b3}. Extrapolated results for $M\to\infty$ are obtained using the expected dependence of the observable critical values on $M$ due to error scaling \cite{suzuki85}. An exemplary fitting is presented in the inset of Fig.~\ref{temp}. As may be concluded from the resulting three critical lines, the area covered by the disordered phase shrinks with decreasing temperature. To classify the behavior at $T=0$, we find in the same manner the critical line in the $U$-$T$ plane for $\mu/U = 0.5$, as presented in Fig.~\ref{temp}. The critical value of $U/J$ at $T\to0$ approaches $\sim8$, which is finite, thus we expect a quantum phase transition to occur in the system. By expanding the basis to include $n=3$, we obtain a phase diagram for a broader range, $0 \leq \mu/U \leq 2$, depicted in Fig.~\ref{b4}. We find a lobe-like behavior typical for strongly correlated bosonic systems. In the inset, the dependence of dynamical self-correlations $\mathcal{R}_{kk'}$ on $|k-k'|$ at the critical point reveals the dynamical nature of the solution, especially at lower temperatures.

{\it Experimental realization.}---
The all-to-all tunneling, which is the major issue of the studied system, can be realized experimentally in a specific optical lattice formed as an array of atomic traps in a shape of $N$ elongated vertical and horizontal rods in a wood-pile arrangement (see Fig.~\ref{b4}). Every horizontal (vertical) rod of a condensate is linked via a Josephson junction \cite{cataliotti} to each of perpendicular counterparts, so that the number of nearest neighbors of a given rod is $z = N$, implying that the system is {\it fully connected}. The corresponding Bose-Hubbard Hamiltonian \eqref{ham} is written in terms of operators for the $\ell$-th vertical/horizontal ($v/h$) rod of condensate at position $\bm{R}_\ell$, related to the second-quantized total condensate wave function $\varPsi(\bm{r})=\sum_\ell[a_{v\ell}\varphi_{v\ell}(\bm{r})+a_{h\ell}\varphi_{h\ell}(\bm{r})]/{N_b}$
where $N_b$ is the number of bosons in the system and $\varphi_{\alpha\ell}(\bm{r})$ is the wave function of the $i\equiv\alpha\ell$-th rod, $\alpha=v,h$. In terms of $\varphi_{\alpha\ell}(\bm{r})$, the first term in Eq.~\eqref{ham} contains the Josephson amplitude $J_{\ell\ell'}=\frac{\hbar^2}{2m}\int d^3\bm{r}\nabla\varphi_{v\ell}(\bm{r})\cdot\nabla\varphi_{h\ell'}(\bm{r})+\int d^3\bm{r} \varphi_{v\ell}(\bm{r})V(\bm{r})\varphi_{h\ell'}(\bm{r}) $ and describes the tunneling of bosons between condensates. $V(\bm{r})$ is the optical-lattice trapping potential, precise form of which is unimportant here, since it is implicit in $J_{\ell\ell'}$. Note, that the translational symmetry makes $J_{\ell\ell'}$ rod-independent, $J_{\ell\ell'}\equiv J$. Furthermore, $U=(2\pi\hbar^2 l_s)/(mN)\sum_{\alpha\ell}\int d^3{\bm{r}}|\varphi_{\alpha\ell}(\bm{r})|^4$ quantifies the on-site interaction energy, with $l_s$ being the scattering length of the atoms of mass $m$. Finally, $\mu =\Omega/(2N)\sum_{\alpha\ell} |\bm{R}_\ell|^2$ describes the mean trapping potential, where $\Omega=m\omega^2/2$ with the trap frequency $\omega$.

The distances between parallel rods are randomly distributed around same mean value $l$. In the presence of an artificial gauge potential $\bm{A}$ \cite{dalibard}, the tunneling parameters acquire the Peierls phase $J_{\ell\ell'}\to J_{\ell\ell'}e^{i\theta_{\ell\ell'}};~\theta_{\ell\ell'}=\cramped{(2\pi/\varPhi_0)\int_{\bm{R}_\ell}^{\bm{R}_{\ell'}}\!\bm{A}\mathrm{d}\bm{l}}$ ($\varPhi_0$ is the elementary flux quantum). Gauge potential combined with distance randomness generates variations of $\theta_{\ell\ell'}$ allowing for random frustration of $J_{\ell\ell'}$. For large fluxes over the elementary plaquette ${\cal P}$ of the array, $\varPhi=\sum_{\cal P}\theta_{\ell\ell'} \gg \varPhi_0$,
the phases $\theta_{\ell\ell'}$ randomize and fill the interval $(0,2\pi]$ uniformly. In this limit, the density of eigenvalues of the random matrix $J_{\ell\ell'}/\sqrt{N}$ with increasing $N$ approaches the Wigner semicircular law for Gaussian-distributed $J_{\ell\ell'}$, as in the SK model.
One-dimensional (1-D) quantum gases with extreme aspect ratios ($\sim250:1$) have been created recently in a geometry that makes it possible to study many copies of the 1-D system at the same time. The spacing between the rods was such that the traps were not perfectly isolated, but were coupled by a tunneling matrix element \cite{moritz}.

Another implementation of an infinite-range hopping model on an optical lattice was proposed in Ref.~\cite{danshita}. The idea is to utilize photo-association lasers \cite{jones} that couple all the combinations of two atomic bands with molecular states. The effective Hamiltonian of such system is similar to
that of the Sachdev-Ye-Kitaev model \cite{sachdevye,kitaev} comprising an all-to-all two-body hopping. However, as it was noted, the realization of the envisaged scheme might be still difficult with the current experimental technology.

\begin{figure}[tbp!]
	\includegraphics[width=\columnwidth]{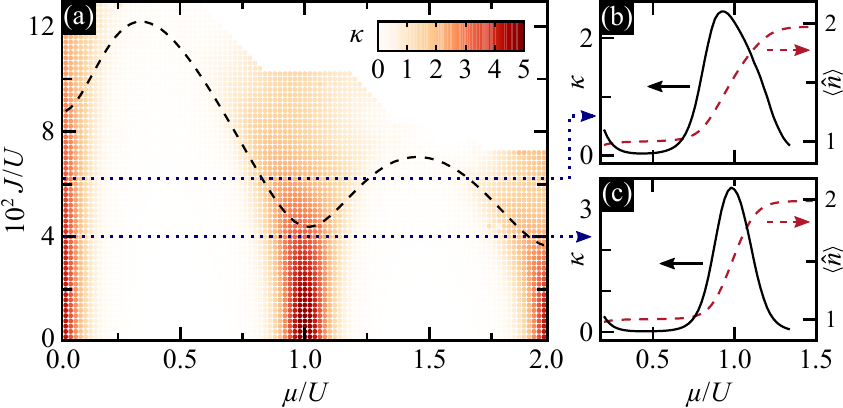}\upcapt
	\caption{(a) Color map of compressibility $\kappa$ (in units of $U$) at $kT/U = 0.05$. Dashed line marks the critical boundary. Results for $\mu/U>1.35$ are obtained for a 5-state basis to account for higher $\avg{\hat{n}}$. (b,~c) Bosonic filling factor $\avg{\hat{n}}$ (dashed line, right axis) and compressibility $\kappa$ (solid line, left axis) vs.\ $\mu/U$ for two values of $J/U$ indicated by dotted lines in (a).}
	\label{compr}
\end{figure}
{\it Discussion.}---
In the widely studied case of diagonal disorder, the behavior of the bosonic filling factor $\avg{\hat{n}}$ as a function of $\mu$ reveals a compressible, $\kappa\equiv \partial \avg{\hat{n}}/\partial\mu>0$, Bose-glass state \cite{fisher}.
To compare our findings with existing works, we present in Fig.~\ref{compr} a color map of numerically calculated compressibility $\kappa$ and its two cuts along constant $J/U$ presenting both $\kappa$ and $\avg{\hat{n}}$. The disordered phase is incompressible without much doubt as $\kappa = 0$ everywhere except the vicinity of integer $\mu/U$ [see Fig.~\hyperref[compr]{\ref*{compr}(c)}] which is an effect of finite temperature (in the $T=0$ phase diagram integer $\mu/U$ do not belong to this phase). This corresponds to the Mott-insulator phase and fully agrees with Ref.~\cite{sengupta}. We also notice that at $J=0$ our system is a pure atomic Mott insulator, so these two phases are in fact the same.
In Ref.~\cite{sengupta}, a new phase was found (called Mott Glass) that shares some of the global properties of the Mott insulator, but locally resembles the Bose glass. Given the described scenario, we conjecture that those results indicate a {\it Griffiths} phase \cite{griffiths}, characterized by rare occurrences of local order in an otherwise disordered medium.

\begin{figure}[tbp!]
	\includegraphics[width=\columnwidth]{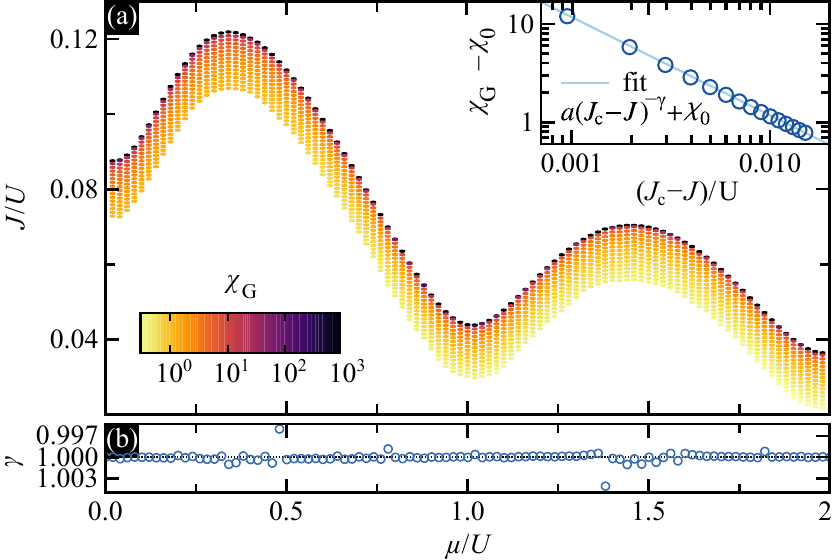}\upcapt
	\caption{(a) Color map of the order-parameter glass susceptibility$\chi_G$ near the glass boundary from Fig.~\ref{b4}. Inset: an exemplary fit of the power-law critical behavior. (b) Universal critical exponent $\gamma$ of $\chi_G$.}
	\label{susc}
\end{figure}
Working with finite $M$, we are confined effectively to nonzero temperature,
and it is difficult to analyse the structure
of the phases based on $\kappa$ alone, since it does not vanish at $T\neq 0$. Hence, $\kappa$ cannot be used to distinguish between different phases, contrarily to ${\mathcal{Q}}_{EA}$.
Thus, we study the order-parameter glass susceptibility $\chi_{\mathrm{G}} = \cramped{\sum _{ij} \Lr{ \abs{ \langle \hat{a}_i \rangle }^2 \abs{ \langle \hat{a}_j \rangle}^2 }_J} / N$, presented in Fig.~\hyperref[susc]{\ref*{susc}(a)}. We find that $\chi_{\mathrm{G}}$ diverges according to the power law $\chi_{\mathrm{G}}\sim\cramped{(J_{\mathrm{c}}-J)^{-\gamma}}$ with the universal exponent $\gamma=1$ in the full range of $\mu/U$, as shown in Fig.~\hyperref[susc]{\ref*{susc}(b)}.

We expect that our findings are robust with respect to the tunneling range, {\it i.e.}, for the short-range-interaction glass problem, we expect qualitatively similar phase diagrams albeit with different numerical values.
This is substantiated by the observation of the behavior of the quantum SG on the Bethe lattice, where the connectivity parameter $z$
can be varied \cite{kopecusadel}. For distributions of hoppings with nonzero mean we envisage the appearance of the superfluid phase,
as in the pure Bose-Hubbard model, possibly coexisting with the glass order depending on the interplay of the model parameters. 
Finally, given the ubiquitous nature of disorder in physical systems, it remains to figure out, 
{\it e.g.}, how the novel but poorly understood topic of many-body localization \cite{mbl}, present also in bosonic systems \cite{mblb}, is related to the issues of quantum-glass transition \cite{burin} of interacting bosons.

\begin{acknowledgments}
{\it Acnowledgments}---Calculations have been carried out using resources provided by Wroclaw Centre for Networking and Supercomputing (\url{http://wcss.pl}), grant No. 449. 
\end{acknowledgments}

\end{document}